\documentstyle[preprint,aps,prl]{revtex}

\begin{document}

\draft
\title{ Four-Particle Anyon Exciton: Boson Approximation}
\author{M. E. Portnoi\cite{byline}  and E. I. Rashba\cite{byline1} }
\address{
Department of Physics, University of Utah, Salt Lake City, UT 84112}

\date{\today}

\maketitle


\begin{abstract}
A theory of anyon excitons consisting of a valence hole and three
quasielectrons with electric charges $(-e/3)$ is presented.
A full symmetry
classification of the $k=0$ states is given, where $k$ is the exciton
momentum. The energy levels of these states are expressed by quadratures
of confluent hypergeometric functions. It is shown that the angular
momentum $L$ of the exciton ground state depends on the distance
between electron and hole confinement planes and takes the values $L=3n$,
where $n$ is an integer. With increasing $k$ the electron density shows
a spectacular splitting on bundles. At first a single anyon splits
off of the two-anyon core, and finally all anyons become separated.
\end{abstract}

\pacs{  }


\section{Introduction}
\label{sec:intro}

Recent progress in the intrinsic interband spectroscopy
\cite{PinDPW93,THWFCRHF90,GBDDNCSHF92,HPDDPW94} of the Fractional
Quantum Hall Effect (FQHE) \cite{TSG82}  made a demand for developing  a
theory of optical phenomena
 accounting in an appropriate way the basic properties
of FQHE phases. Such a theory has to reflect two essential facts:
i) an exciton nature of the low-energy states in a strong magnetic field,
and ii) the presence of an Incompressible Quantum
Liquid (IQL) \cite{Laug83} underlying the FQHE. Therefore, the theory
should describe an exciton against a background of an IQL. There are two
approaches to this problem. The first one is based on exact numerical
treatment of few-electron systems in a spherical geometry
\cite{MacDR90,AR-three,AR-split,MRK,CQ-PR,ZB}. The second one is based
on the anyon exciton (AE) model \cite{RP93} which considers
an exciton as a neutral entity
consisting of a valence hole and several quasielectrons (QE's) carrying
fractional negative charges. These both approaches have a restricted
applicability depending on the value of the parameter $h/l$, here $h$
is the separation between electron and hole confinement planes and
$l=(c\hbar/eH)^{1/2}$ is the magnetic length. Reliable calculations in a
 spherical geometry are accessible only for $h/l\alt 1.5$. On the contrary,
the AE approach fails when $h/l$ is small, but is exact in
the opposite limit $h\gg l$. Indeed, by electrostatic arguments the
size of an AE ground state is about $h$,
 and the model is applicable when the mean
distance between anyons is large as compared to the anyon size which is
about $l$; in the same limit the perturbation exerted by a hole is small.
There is our believe that both approaches
 will result in a similar picture in the
intermediate region $h/l\sim 1$ and, therefore, taken together they
 will provide  a description
of excitons in the whole region of the $h/l$ values.
Experimental data on excitons are not avalable for the region $h>l$ yet,
 and therefore a comparison with the experiment can not
be performed at present. For this reason we abstain from  discussing
the involved problem of the transition probabilities \cite{AR-SSC} and
concentrate on the energy spectra and electron density distribution.
 The relation between the results of  the AE approach and of computations for
few-particle systems is discussed in the last Section.

The oversimplified model  of an AE consisting of a hole and
two semions (anyons having charges $(-e/2))$ has been developed by us
previously \cite{RP93}. In that paper a multiple-branch spectrum of
an AE has been established as a basic signature of the charge
fractionalization in IQL's. Recent data \cite{AR-SSC,RAcambr}
 have revealed the appearance of this structure also in few-particle
computations. It is remarkable that several new exciton branches appear at
very moderate values of $h/l\alt 1$. In what follows we present for
the first time a theory for a realistic case of three QE's
 which corresponds to a $\nu=1/3$ IQL.

\section{Theory}
\label{sec:theory}

Let us consider an exciton consisting of a valence hole with a charge
$(+e)$ and three QE's with electrical charges $(-e/3)$ and statistical
charges $\alpha= -1/3$ ($\alpha=0$ for bosons and $\alpha=1$ for fermions).
In the strong $H$ limit, when the Coulomb energy
 $\varepsilon_{C} = e^{2}/\epsilon l \ll \hbar \omega_{c}$,
where $\omega_{c}$ is the cyclotron frequency, it is convenient to use
dimentionless variables scaled in units $\varepsilon_{C}$, $l$ and $e$.
Instead of hole, ${\bf r}_{h}$, and anyon, ${\bf r}_{i}$, coordinates
it is convenient to introduce the following two-dimensional coordinates:
\begin{equation}
{\bf R} = {1\over 2}({\bf r}_{h} + {1\over 3}\sum_{i=1}^{3}{\bf r}_{i})~,~
  {\bbox {\rho}} =  {1\over 3}\sum_{i=1}^{3}{\bf r}_{i} - {\bf r}_{h}~,
{\bf r}_{ij} = {\bf r}_{i} - {\bf r}_{j}, ~{ij}={12},~ {23}, ~{\rm and}~ {31}~.
\label{eq1}
\end{equation}
Complex coordinates $z_{jl}=x_{jl}+iy_{jl}$, as well as ${\bf r}_{jl}$,
 are not independent. Indeed:
\begin{equation}
{\bf r}_{12} + {\bf r}_{23} + {\bf r}_{31} = 0~.
\label{eq2}
\end{equation}
Since AE is a neutral entity it possesses an in-plane
 momentum ${\bf k}$ \cite{GD}.
 As a  basis for the exciton wave functions
 in the Halperin's quasi wave function representation \cite{Halp84,Laug90}
the functions
\begin{equation}
\Psi_{L,{\bf k}} = \exp\{i{\bf k}{\bf R} + i({\rho_{x}}Y-{\rho_{y}}X)
  -(\bbox{\rho} - {\bf d})^{2}/4\}
P_{L}(\dots \bar{z}_{jl} \dots) \prod_{jl} (\bar{z}_{jl})^{\alpha}
\exp\{-|{z_{jl}}|^{2}/36\}~
\label{eq3}
\end{equation}
may be chosen. Here ${\bf d}={\hat{z}}\times {\bf k}$,
 and $(-{\bf d})$ is the exciton dipole
moment. Exponential factors in Eq.~\ref{eq3} ensure the translational
magnetic symmetry and belonging the function to the ground Landau level
 both for anyons and a hole. A function  $P_{L}(\dots \bar{z}_{jl} \dots)$
is a homogeneous polynomial of the degree $L$ which is symmetric in all
coordinates $z_{i}$. The function $\Psi_{L,{\bf k}}$ describes an exciton
with the momentum ${\bf }k$ and the internal angular momentum $(-L)$.

For a three anyon problem polynomials
 $P_{L}$ depend on three variables $\bar{z}_{12}$,~
$\bar{z}_{23}$, and ~$\bar{z}_{31}$. If to take into account the constraint
$\bar{z}_{12} + \bar{z}_{23} + \bar{z}_{31} = 0$, one can show that the total
number of linearly independent symmetric polynomials equals
$[L/6]+1$ for even $L$ and $[(L-3)/6]+1$ for odd $L$. Square brackets
designate the integer part. Odd $L$ polynomials start with $L=3$ because
of Eq.~\ref{eq2}. Even-$L$  polynomials $P_L$ may be classified as:
\begin{equation}
P_{L,M} = \bar{z}_{12}^{L-4M}\bar{z}_{23}^{2M}\bar{z}_{31}^{2M} +
\bar{z}_{23}^{L-4M}\bar{z}_{31}^{2M}\bar{z}_{12}^{2M} +
\bar{z}_{31}^{L-4M}\bar{z}_{12}^{2M}\bar{z}_{23}^{2M} ~,
\label{eq4}
\end{equation}
where $M = 0,~1,\dots [L/6]$. The only symmetric polynomial with $L=3$
is a Vandermonde determinant in the variables $\bar{z}_{jl}$:
\begin{equation}
V = (\bar{z}_{12} - \bar{z}_{23}) (\bar{z}_{23} - \bar{z}_{31})
(\bar{z}_{31} - \bar{z}_{12})~.
\label{eq5}
\end{equation}
As distinct from the polynomials $P_{L,M}$ with even $L$, which are
symmetric both in the bosonic
 permutations $\bar{z}_{1}\leftrightarrow \bar{z}_{2}$
and permutations
$\bar{z}_{12}\leftrightarrow \bar{z}_{23}$, etc., the polynomial
$V$ is symmetric in the permutations of the first and
antisymmetric in the permutations of the second type. All independent
odd-$L$ polynomials can be represented in the form:
\begin{equation}
P_{L,M} = V P_{L-3,M}~,~~P_{3,0} = V~.
\label{eq6}
\end{equation}
\noindent To our best knowledge, in the previous studies only the even-$L$
polynomials have been taken into account \cite{ET92}. When choosing
polynomials $P_{L,M}$ we have not imposed the hard core constraint and
defer the discussion of the related properties to what follows.

In the basis of polynomials $P_{L,M}$ the set of quantum numbers consists
of ${\bf k},~L$ and $M$. The only disadvantage of the functions
$\Psi_{L,M,{\bf k}}$ is that the scalar products
 $<LM{\bf k}|LM'{\bf k}> \neq 0$ for $M\neq M'$. As a result,
the matrix ${\hat B}$ of these scalar products which is diagonal in
${\bf k}$ and $L$ has a block-diagonal form; the block sizes
increase with $L$. The Schr$\ddot{\rm o}$dinger equation is
${\hat H}\chi=\varepsilon {\hat B}\chi$. All integrations in scalar
products and matrix elements of ${\hat  H}$
 should be performed over four variables ${\bf r}_{h},~
{\bf r}_{1},~{\bf r}_{2}$, and ${\bf r}_{3}$. It is convenient to
choose the new variables ${\bf R},~{\bbox{\rho}}$ and three ${\bf r}_{ij}$ and
take into account the constraint of
Eq.~\ref{eq2} by the usual $\delta$-function
transformation. It adds a new variable ${\bf f}$, but all calculation become
 symmetric in the anyon variables. As an example we show here the
 transformation of  a matrix element with polynomials $P_{L,M}$
substituted by monomials
 ${\bar z}_{12}^{n_3}{\bar z}_{23}^{n_1}{\bar z}_{31}^{n_2}$.
 After the $\delta$-function
transformation the matrix element takes the form:
\begin{equation}
<n_{1}n_{2}n_{3}{\bf k}|n_{1}^{'}n_{2}^{'}n_{3}^{'}{\bf k}>
 = \int {d{\bf f}\over {(2\pi)^{2}}}
 {\cal M}^{\alpha}_{n_{1}n_{1}'}({\bf f})
 {\cal M}^{\alpha}_{n_{2}n_{2}'}({\bf f})
 {\cal M}^{\alpha}_{n_{3}n_{3}'}({\bf f})~,
\label{eq7}
\end{equation}
where
\begin{eqnarray}
{\cal M}^{\alpha}_{nn'}({\bf f}) = && 2\pi i^{|n-n'|}~
{\Gamma(\max\{n,n'\} + \alpha + 1)\over {|n-n'|!}}~
2^{(n+n')/2 + \alpha}~ 3^{(n+n') + 2(\alpha + 1)}~ t^{|n-n'|/2} \nonumber \\
&&\times
\exp\{i\varphi_{\bf f}(n'-n)\}~
\Phi(\max\{n,n'\} + \alpha + 1,~ |n - n'| + 1~; -t)~.
\label{eq8}
\end{eqnarray}
Here $t=9f^{2}/2$,~ $\varphi_{\bf f}$ is the azimuth of ${\bf f}$,~ and
$\Phi$ is a confluent hypergeometric function.

Quasi wave functions  of Eq.~\ref{eq3} differ from bosonic functions
only by a statistical factor $\prod_{jl} (\bar{z}_{jl})^{-\alpha}$.
It is of most importance for small values of $|z_{ij}|\approx l$. For
the same distances the deviation of the anyon-anyon interaction law from
a Coulomb law  \cite{Morf} also should be taken into account and results in
comparable contributions. In what follows we restrict ourselves with
a Coulomb interaction in the Hamiltonian ${\hat H} = V_{aa} + V_{ah}$~,
where $V_{aa} = (1/9)\sum_{ij} r_{ij}^{-1}$ and
$V_{ah} =  - (1/3)\sum_{i}|{\bf r}_{hi}+h{\hat z}|^{-1}$~,
and omit the statistical factor in $\Psi_{L,M,{\bf k}}$.
  This last
approximation strongly simplifies all calculations. Indeed, with
$\alpha=0$ all functions $\Phi$, Eq.~\ref{eq8}, turn into polynomials
if one performs the Kummer transformation,
 $\Phi(\beta,~\gamma;~t) = e^{t}\Phi(\gamma - \beta,~\gamma;~-t)$
and takes into account the fact that $\gamma - \beta $ are non-positive
integers for all functions (\ref{eq8}).  As a result, integrals
(\ref{eq8}) can be calculated exactly. The matrix elements of $V_{aa}$
 differ from those of Eq.~\ref{eq7} by changing in one of the
${\cal M}^{\alpha}_{nn'}$ functions $\alpha$ for $(\alpha -{1/2})$.
This latter function can not be reduced to a polynomial, hence, one-fold
integration over $f$ should be performed numerically. The most complicated
are the matrix elements of $V_{ah}$~. To simplify them it is convenient
to perform a Fourier transformation
$V_{ah}({\bf r})=
\int {d{\bf q}V_{ah}({\bf q})\exp(i{\bf q}{\bf r})}/(2\pi)^{2}$~.
In this representation the integrand contains again three factors
${\cal M}_{n_{i}n'_{i}}^{\alpha}$~.
 One of them has ${\bf f}$ as an argument, while
two others have $({\bf f}\pm {\bf q}/3)$~. For $\alpha=0$ the
integrations over $f$ and over the angle between ${\bf f}$ and ${\bf q}$ may
be performed. We choose $\Psi_{L,M,{\bf k}}$ in such a way that
the matrix elements of ${\hat H}$ are real, and the diagonal elements of
${\hat B}$ equal unity.
 The final expression for the matrix element of $V_{ah}$ is:
\begin{equation}
<LM{\bf k}|V_{ah}|L'M'{\bf k}>= - \int_{0}^{\infty}\exp(-3q^{2}/2-qh)
J_{|L-L'|}(kq)Q_{LM,L'M'}(q) dq~,
\label{eq9}
\end{equation}
where $J_{|L-L'|}(kq)$ are Bessel functions, and the
functions $Q_{LM,L'M'}(q)$, real and symmetric in the $LM$ indices, are
polynomials in $q$. If ${\bf k}=0$, the integral equals zero
for $L\neq L'$ as it follows from the angular momentum
conservation law. Therefore, for ${\bf k}=0$ the total Hamiltonian has
a block-diagonal form. The lower polynomials are of a simple form:
$Q_{00,00}=1,~Q_{20,00}=q^{2}/2,~Q_{20,20}=1-q^{2}+q^{4}/4$. Since
the blocks are of the $1\times 1$ size for the
even-$L$ polynomials with $L<6$ and the odd-$L$ polynomials with $L<9$,
the corresponding wave
functions do not depend on $h$, and the energies may be expressed in terms of
quadratures from transcendental and elementary functions.
 If $L$ is even and $6\leq L<12$,
the coefficients of $2\times 2$ secular equations may be expressed in
the same terms, etc.   With increasing $h$ the anyon-hole
attraction becomes smaller, the effect being stronger the less is $L$.
The block-diagonal structure of ${\hat H}$ allows the energy level
 intersections for $k=0$, for $k\neq 0$ these level crossings turn
into anticrossings.

In Fig.~\ref{f1}  the energy spectrum $\varepsilon(k)$ is shown for two values
of $h$. The following regularities are distinctly seen. With increasing
$h$, the levels with higher $L$ values draw closer to the spectrum bottom.
The branches with small $L$ values have a larger dispersion. The
anticrossings are wider the larger the $k$ values and the lesser the
differences $|L-L'|$ are. It is remarkable that  the
 branches with a negative dispersion near $k=0$ exist; it may appear
 even in the ground state, Fig.~\ref{f1}a. For a two-semion problem
the dispersion near ${\bf k} = 0$ is always positive \cite{RP93}.

The most interesting property of AE's which follows from our calculations
is the distribution of the anyon density $D_{l}({\bf r}, {\bf k})$
around the hole
\begin{eqnarray}
D_{l}({\bf r}, {\bf k}) = && {1\over{2\pi}} \sum_{LM,L'M'}
 \cos[(L-L')\theta] \nonumber \\ &&\times
\int_{0}^{\infty}{ dq~q~\exp(-3q^{2}/2)~J_{|L-L'|}(q|{\bf r}-{\bf d}|)}~
Q_{LM,L'M'}(q)~ \chi_{L'M'}^{l~ \ast}({\bf k}) \chi_{LM}^{l}({\bf k})~,
\label{eq10}
\end{eqnarray}
 where $l$ numerates spectrum branches, and $\theta$ is the angle between
the vectors ${\bf d} - {\bf r}$ and ${\bf d}$, ${\bf d}$ is the
center-of-mass of the anyon density.

\section{Zero momentum anyon excitons}
\label{sec: zero}

For ${\bf k}=0$ the density distribution $D(r,0)$ is shown in Fig.~\ref{f2}
 for  the states with $L\leq 6$.
 It has been mentioned above that $D_{L}$ do not depend on $h$ for
 $L\leq 5$. The state $L=3$ is remarkable in some
sense. It is the first to show a density minimum at $r=0$, and it
generates all even-$L$ polynomials, Eq.~\ref{eq6}. For $L=6$
 there are two eigenfunctions; they depend on $h$. In Fig.~\ref{f2} they
are shown for $h=0$; the lower energy component is drawn by a solid line.
 With increasing $L$ the functions
$D_{L,M}$ become broader, and as a result some of the states with large
$L$ values draw closer to the spectrum bottom
when $h$ is growing,  Fig.~\ref{f1}.
However, it is seen from Fig.~\ref{f3} that only the states with
 $L=3n$,~$n\geq0$ are integers, reach the spectrum bottom.
The bottom states described by even- and odd-$L$ polynomials
alternate. The periodicity with the period $\Delta L=3$ resembles the
Laughlin's result \cite{LaugPR} for a three-electron system.
 We attribute this periodicity
 in $L$ to the quantization rule for identical particles, since the
 rotation angle between two exchange points (which enters into
the semiclassical
quantization rule \cite{Sivan}) equals $2\pi/3$.

The functions with $L=6M$,~$M\geq 1$ are integers, depend on $h$ strongly.
For $L=6$ the electron density distribution $D(r,0)$ changes rapidly
in the vicinity of $h=2$, Fig.~\ref{f4}.  For $h<2$ the low energy
  component of the $L=6$ state is close to $\Psi_{6,0}$
(for $h=0$ the overlap is 0.96), while for $h>2$ it approaches
the hard core state $\Psi_{6,1}$ (for $h=3$ the overlap is 0.97).
The function $\Psi_{6,1}$ is the first in the series of the even-$L$
polynomials
$P_{6M,M} = ({\bar z}_{12}{\bar z}_{23}{\bar z}_{32})^{2M},~ M\geq 1$,~
forming ground states for the large $h$ values. The density $D_{L,M}(r,0),
{}~ L=6M$, has a single maximum  for each of these states.
The maximum of the density $\tilde{D}_{L,M}(r,0)$ found after averaging
over ${\bf r}_{h}$ is determined by a simple equation $r_{L} =  \sqrt{2L}$;
the maximum of $D_{L,M}(r,0)$ is pretty close to this point.
 In the vicinity of the maximum of $D_{L,M}(r,0)$
 the three particle correlation function
 reaches the maximum for an equilateral triangle. It is
just the result which is expected in the $L\gg 1$ limit from electrostatic
arguments. The value of  $r_{ij}^{2}$, found both as a mean value in
a $\Psi_{6M,M}$ state and by the semiclassical approach
in terms of the distances between anyon guiding centers
 \cite{Sivan}, satisfies the equation $L = 6M = r_{12}^{2}/6 - 2$.

The density distribution for $\Psi_{3,0}(r)$ shows a single maximum,
Fig. 2. In the vicinity of it the three-particle
 correlation function reaches the
maximum for an equilateral triangle configuration of anyons with
 $r_{12}^{2} \approx 18$. In this respect the functions  $P_{6,1}$ and
$P_{3,0}$
show a similar behavior.  A strong difference between them is reflected
 in the anyon two-particle correlation functions $W(r)$:
\begin{equation}
W_{6,1} = {r^{2}\over 192\pi}(1+{1\over 2}(r/6)^{4})\exp(-r^{2}/12),~
W_{3,0} = {1\over 48\pi}(1+{1\over 6}(r/2)^{4}) \exp(-r^{2}/12).
\label{eq11}
\end{equation}
They are shown in Fig.~\ref{f5} . The function $W_{6,1}$ has
a hard core behavior, while $W_{3,0}$ reaches its absolute maximum
at $r=0$ and has the second maximum which is only by a factor 0.97 lower than
the main one. Of course, all bottom state odd-$L$
  polynomials with $L\geq 9$ show
a hard core behavior.
It follows from the above data that with increasing $h$ the charge
fractionalization
results in the broadening of the anyon density distribution
 in the low-energy $k=0$ states. The dependence of the exciton binding
energy on $h$ is close to the Coulomb law, but the numerator is
considerably less than unity. In Fig.~\ref{f3}  the energy of an ordinary $k=0$
magnetoexciton, $\varepsilon_{ME}(h)$, is shown for comparison.
Because of the existence of a hidden symmetry,
  $\varepsilon_{ME}(h)$ gives an
exact result in the $h \rightarrow 0$ limit. The AE model
becomes exact in the opposite limit, $h \rightarrow \infty$. The
both curves should be matched somewhere in the intermediate region,
$h \sim 1$.

\section{Momentum dependence of the exciton charge density}
\label{sec: momentum}

Since the exciton dipole moment ${\bf d}$ differs from ${\bf k}$ only by
the factor $el^2$ and the rotation by $\pi/2$, Sect.~\ref{sec:theory}, one
can expect that with increasing $k$ the electron density splits into
bundles, their charges being multiples of $e/3$.
The splitting of the electron shell into two well separated quasiparticles
has been observed previously for a two-semion exciton \cite{RP93}. For a
three-anyon exciton the pattern are much more impressive. They are shown
in Fig.~\ref{f6}  for $h=3$ when the criterion of the large   electron-hole
separation is fulfilled. The distribution which is cylindrically symmetric
for $k=0$ transforms with increasing $k$ into a distribution with a
single split off anyon, $k=2$ and 3. Two anyons constituting the
exciton core show a slight but distinct splitting in a perpendicular
direction. This core changes its shape with $k$, but remains stable
for a long. Finally, for rather large $k$ values, it  splits in the
 ${\bf d}$ direction as it is seen in the last figure, $k=6$. In all
figures the center-of-mass of the electron density is at the point
$x=k$, and the asymmetric distribution of the density arises due to
the odd-$L$ polynomials. In the absence of them the distribution had
to be symmetric with respect to the point ${\bf d}$.

The well outlined profiles of the electron density seen in Fig.~\ref{f6}
  may be somewhat smeared  by the oscillatory screening
inherent in IQL's \cite{screening}. Nevertheless, the basic pattern of the
charge separation in an exciton should strongly influence the $k$
dependence of the magnetoroton-assisted recombination processes since
charge density excitations are left in a crystal afterwards.

\section{Discussion}
\label{sec:discussion}

The above theory is based on a Coulomb interaction between all the particles
and neglecting the statistical factor
$\prod_{jl} (\bar{z}_{jl})^{\alpha}$ in Eq.~\ref{eq3}.
For small $L$ values the theory results in dense states
 which can not be described by the
AE model, and for larger $L$ values in
 more loose states to which the AE model should be applicable.
The $L_3$ state or, more probable, the $L_6$ state are the first candidates.
If one takes into account the statistical factor, the effective repulsion
of anyons should increase. Therefore, we expect that the critical value
of $h$, $h_{\rm cr}$, when the loose states reach the spectrum bottom must
decrease. According to Fig.~\ref{f3},
 $h_{\rm cr}\approx 2$ in the AE model, while
the computations in the spherical geometry  result in
$h_{\rm cr}\approx 1$ for $\nu=1/3$ \cite{AR-SSC,RAcambr}.
These data are in a reasonable agreement. The more intriguing question is
whether the realistic anyon form-factors and quasipotentials may result
in some new types of the $k=0$ ground states, e.g., the two-scale states like
the low-energy component of the
 $L=6$ state in Fig.~\ref{f4}, $h=1$.  It has been proposed by
Apalkov and Rashba \cite{AR-split},
 that the two-scale exciton observed by
them in computations for a $\nu=1/3$ IQL consisted of a two-anyon core and a
split off anyon (it resembles the exciton shown in our Fig.~\ref{f6},
$k=2$, if described in the $(L,k)$ rather than ${\bf k}$ representation).
 Such a configuration may be energetically favorable for a
$k=0$ exciton if the two-QE repulsion at small distances is suppressed well
below its Coulomb value. This hypothesis is in agreement with the data
of  B\'{e}ran and  Morf \cite{Morf}. Therefore, the configurations of
anyon shells of AE's with moderate values of $h$
 are expected to be sensitive to the QE interaction law, and
calculations based on realistic quasipotentials are desirable.

\section{Acknowledgments}
\label{acknow}

We are grateful to Prof. A. L. Efros for numerous suggestive discussions
and to E. V. Tsiper for valuable comments related to odd-$L$ polynomials.
The support by Subagreement No. KK3017 from QUEST of UCSB is acknowledged.

\begin{figure}
 \caption{ Anyon exciton dispersion law $\varepsilon(k)$
 for two values of $h$.
Mention the opposite signs of the ground state  dispersion
for both curves. Anticrossings become tiny with increasing $h$.
 Numbers show $L$ values. For more detail see text.
\label{f1} }
\end{figure}

\begin{figure}
 \caption{ Axisymmetric electron density distributions $D_{L}(r,0)$
for the states with $L\leq 6$. Two $L=6$ states are shown for
$h=0$; the lowest state is shown by a solid line. Numbers show $L$ values.
\label{f2} }
\end{figure}

\begin{figure}
 \caption{
 The energy $\varepsilon(0)$ and the
 electron density $D_{L}(0,0)$ at the point
${\bf r}_{h}$ where the hole resides plotted {\it vs} $h$
for the ground state of an exciton with $k=0$. The ground state energy of an
anyon exciton is shown by a solid line; the dots on it show the
positions of the intersections between the energy levels with different
 $L$-values. For comparison the energy of a conventional magnetoexciton
$\varepsilon_{ME}$ with $k=0$
is shown by a dashed line. Numbers near the $D_{L}(0,0)$ curve show
the $L$ values.  Only
the states with $L=3n$ reach the spectrum bottom (as an exclusion the
state $L=2$ appears as a bottom state in an extremely narrow region
of the $h$ values).
\label{f3} }
\end{figure}

\begin{figure}
 \caption{Electron density distribution $D(r,0)$ in the $L=6$ states
with $k=0$
for three values of $h$. The density in the lower energy state is
 shown by solid a line. Consecutive numbers, $l$, of the energy
levels are shown near the curves.
\label{f4} }
\end{figure}

\begin{figure}
 \caption{Normalized anyon pair correlation function $W(r)$ for the states
$\Psi_{6,1}$ and $\Psi_{3,0}$; $k=0$.
\label{f5}}
\end{figure}

\begin{figure}
 \caption{Electron density distribution in an anyon exciton for
different values of $k$. A hole is in the origin, the $x$ axis is chosen
 in the ${\bf d}$ direction. The center of the electron density distribution
is at $x=k$.
\label{f6}}
\end{figure}
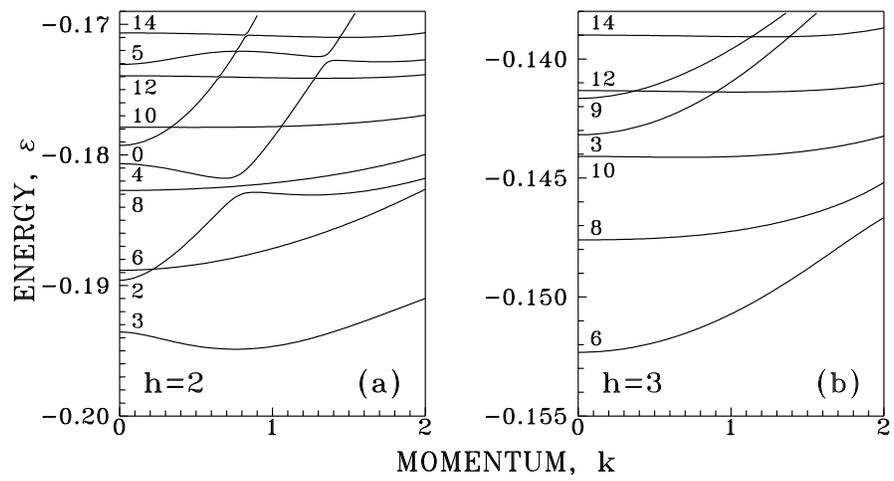
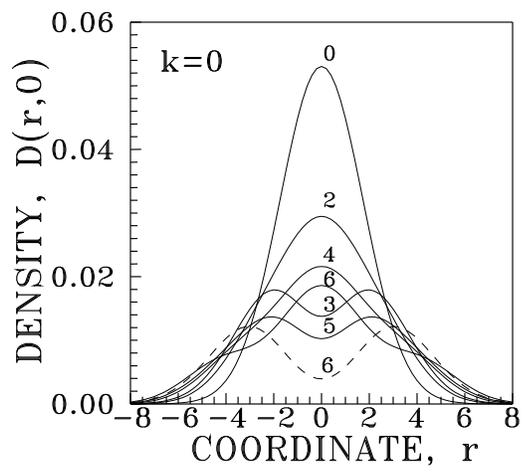
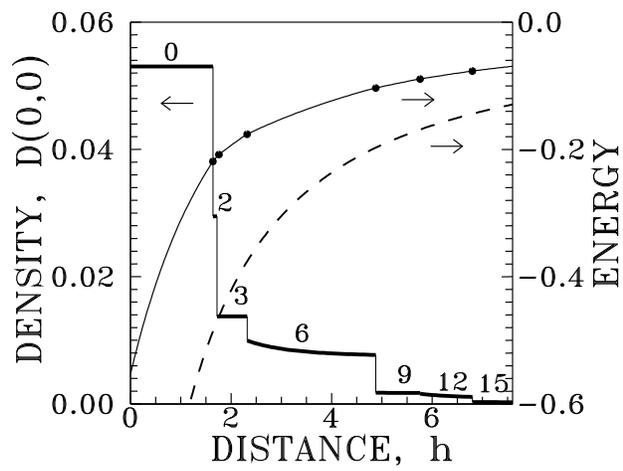
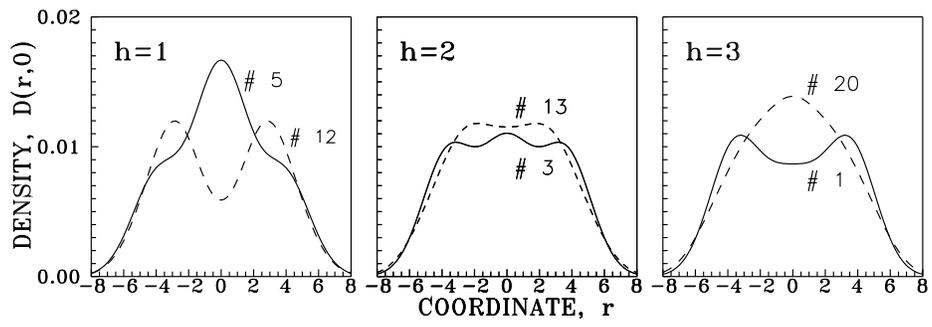
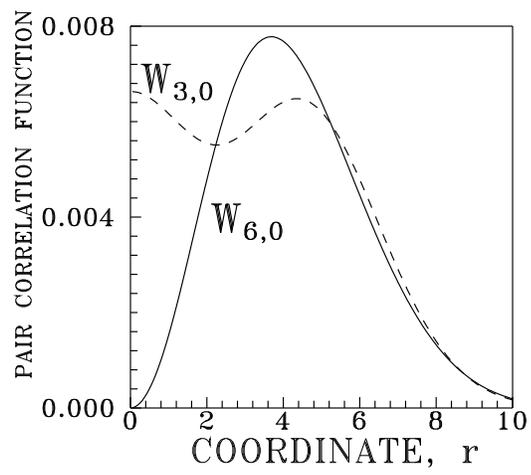
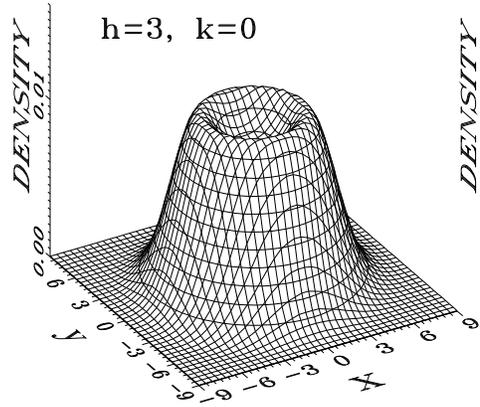
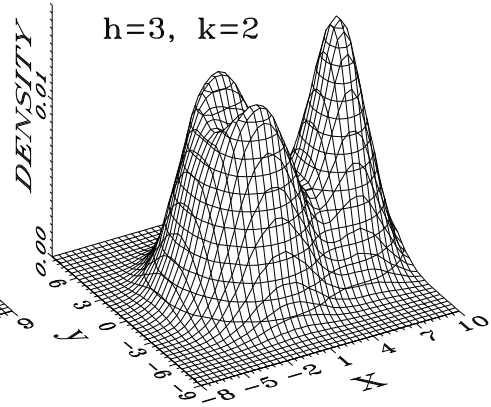
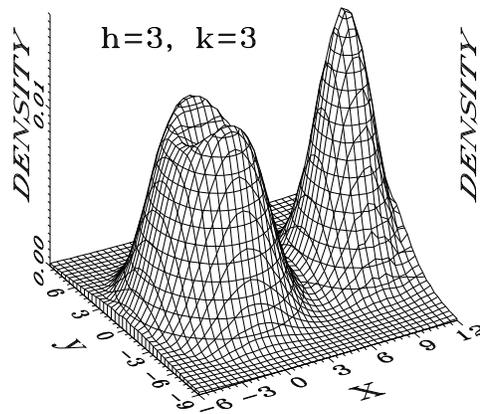
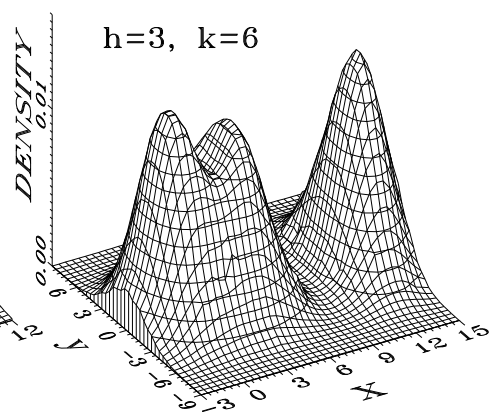
\end{document}